\documentclass[a4paper,11pt]{article}
\pdfoutput=1 

\usepackage{jheppub} 
\usepackage{physics}
\usepackage[T1]{fontenc} 

\begin{document}

\title{String junctions suspended between giants}

\author{David Berenstein,}
\author{Adolfo Holguin}
\affiliation { Department of Physics, University of California at Santa Barbara, CA 93106
}

\emailAdd{dberens@physics.ucsb.edu}
\emailAdd{adolfoholguin@physics.ucsb.edu}
\abstract{
We construct $(p,q)$ string junction solutions suspended between both sphere and AdS giant  gravitons in $AdS_5\times S^5$. Our results extend easily to more general half BPS geometries of LLM type.
These carry angular momentum in the directions of the worldvolume of the giant gravitons. We argue that these are charged under a central extension of the supersymmetry algebra similar
to the one that has appeared in the works of Beisert for the ${\cal N}=4 $ spin chain. We also argue that they are BPS with respect to this central extension.
We show that apart from some kinematical details, the junctions end up solving the same minimization problem that appears in the Coulomb branch of ${\cal N}=4 $ SYM. Their mass and shape is independent of the angular
 momentum $Q$ that the junction carries.
}

\maketitle

\flushbottom
\section{Introduction}
Integrability and supersymmetry are powerful tools for studying non-perturbative phenomena of a very special class of large $N$ gauge theories. The most studied example of this is the appearance of integrable structures in the $\mathcal{N}=4$ super Yang-Mills theory, where the dilatation operator acting on single trace operators behaves like the Hamiltonian of an integrable spin chain; for a review see \cite{Beisert:2010jr}. An important lesson from this analysis shows that the spectrum of anomalous dimensions is controlled by a \textit{central extension} of the superconformal algebra which describes the quasi-momenta of the defects on a long-string \cite{Beisert:2005tm}. In the spectrum of single trace operators of the $\mathcal{N}=4$ theory, the total central charge vanishes. However, the central charge still controls the individual excitations of the spin chain.
 The supersymmetry algebra fixes the dispersion relation of the worldsheet defects to be of the form:
\begin{equation}
\begin{aligned}
 \Delta- J&= \sum_{i=1}^M \sqrt{1+ f^2(\lambda) \sin^2\left(\frac{1}{2}p_i\right)}.\\
 \sum_{i=1}^M p_i&=0
\end{aligned}
\end{equation}
In this case $\Delta$ is the dimension of the spin chain operator, $J$ is the energy of the ferromagnetic ground state, and we take $\Delta,J \to \infty$, keeping $\Delta-J$ finite
to obtain the infinitely long spin chain. The quantities $p_i$ are the quasimomentum of the excitations.
At strong coupling, the quantity $\sin\left(p_i/2\right)$ describes the geometric length of each giant magnon on the closed string \cite{Hofman:2006xt}, including motion effects, while $f(\lambda)= \frac{\lambda}{4 \pi}$ corresponds to the string tension in $AdS$ units.  Closed strings states do not carry any central charge due to the level matching condition (this is due to the cyclic property of the trace \cite{Berenstein:2002jq}).

In order to have states that are charged under this central extension, one has to consider sources for the central charge. In this case, the sources are supersymmetric D-branes. They are realized by giant gravitons \cite{McGreevy:2000cw}. The ground state for an open string stretching between giant gravitons is expected to have a similar dispersion relation \cite{Berenstein:2014zxa},
\begin{equation}
    \Delta-J = \sqrt{Q^2 + \frac{\lambda^2}{16 \pi ^2}|\xi-\Tilde{\xi}|^2},
\end{equation}
where $Q$ is one of the other angular momenta along either $S^5$ or $S^3\subset AdS_5$, and $\xi, \Tilde{\xi}$ are the geometric positions of of the string's end points in a special set of coordinates. These are related to the LLM coordinates  \cite{Lin:2004nb} of the giant gravitons.

These position differences evaluate to the central charge of the state. In some cases this can be related directly to the Higgs mechanism in ${\cal N}=4$ SYM in the Coulomb branch, where the central charge is carried by the W bosons. This allows one to read $f(\lambda)$ from field theory. These W bosons are BPS states. The central charge is necessary for these to be protected by supersymmetry and to form multiplets with spin less than or equal tot one. This dispersion relation for the ground state of open strings has been verified directly up to three loop order in the works \cite{Berenstein:2013eya, Berenstein:2014isa, Dzienkowski:2015zba}. The state can also be understood as a giant magnon that has been cut to end on the giant gravitons. A similar dispersion relation with $Q$ can also be found for bound states of giant magnons on the spin chain side \cite{Dorey:2006dq} and rotating  magnon solutions can be found in supergravity \cite{Spradlin:2006wk}. The point is that these states are very well understood both in the
${\cal N}=4 $ SYM field theory and in the supergravity setup. They can be matched to one another and they are consistent with each other. The origin of the central charge in the spin chain can be traced to the the central charge of the Coulomb branch, Similarly, the central charge extension in the gravity side can be related to the central charge extension
of flat space time, where the central charge is carried by the separation between two D-branes. The difference is that in flat space, states with non-zero central charge are allowed.
In $AdS$, the ${\cal N}=4$ superconformal algebra does not allow for a central extension, so ultimately all states are neutral under the central charge. Their
individual parts can be charged and protected under a central charge that arises as an effective symmetry on the worldsheet of the string when it is allowed to end someplace. This requires ignoring back reaction on the endpoints. These are D-branes with mass of order $N$.  Therefore $N$ must be very large and $1/N$ corrections to the D-brane motions can be ignored.

Type IIB superstrings and $\mathcal{N}=4$ super Yang-Mills are known to admit many more BPS bound states related to each other by $SL(2,\mathbf{Z})$ duality, such as $(p,q)$ strings and dyons \cite{Sen:1994yi,Sen:1997xi,Callan:1997kz,Dasgupta:1997pu}. By exchanging duality frames, it is natural to expect that an open $(p,q)$-string has a similar ground state energy, with the fundamental string tension $T_{(1,0)}= \frac{\lambda}{4 \pi}$ being replaced by the tension of a BPS $(p,q)$-string, $T_{(p,q)}= T_{(0,1)} |p+ \tau q|$. These types of arguments based on S-duality were used to argue also that $f(\lambda)$ receives no corrections \cite{Berenstein:2009qd}, before the precise connection to the Coulomb branch of ${\cal N}=4 $ SYM realized by sphere giant gravitons and  AdS giant gravitons \cite{Grisaru:2000zn,Hashimoto:2000zp} (the latter are also called dual giant gravitons) was fully understood.

Similarly, the ground state energies of more complicated BPS bound states in $AdS_5\times S^5$, such as string junctions, should be controlled by an appropriate notion of central charge extension. In flat space, the junctions are protected by supersymmetry \cite{Callan:1997kz,Dasgupta:1997pu}, they can also be found directly in  $SU(3)$ gauge theory as dyons that bind monopoles and W bosons \cite{Bergman:1997yw} and they carry a central charge. They are only $1/4$-BPS states in the Coulomb branch, unlike the half BPS W-bosons.

 In this note we argue that these states also exist in the dual supergravity theory and that they are protected by supersymmetry in the same way that giant magnons are.
 We do this by finding open string  junction configurations in $AdS_5\times S^5$ whose energy saturates the BPS bound for the centrally extended supersymmetry algebra that one expects
 in the presence of sphere giant gravitons and AdS giant gravitons. The physics of such string junctions is a minimal modification of that of flat-space. Also, there are new lines of marginal stability that arise because the string segments are forbidden from crossing certain regions of the geometry.
 The junctions we study are different than the ones studied in \cite{Sfetsos:2007nd}, which are attached to external  sources on the boundary of AdS.

 The paper is organized as follows. in section \ref{sec:seg} we discuss the energies of  the ground states of open string segments for $(p,q)$ strings suspended between two endpoints. These carry additional angular momentum and  saturate a BPS condition.
 In section \ref{sec:junctions} we discuss how to assemble these segments into an energy of a junction. The problem of finding the ground state is converted to a minimization problem where one varies the position of the junction and the  fraction of angular momentum carried by each segment.
 We show that the problem can be reduced in the end to the same minimization problem for string junctions in flat space. The upshot is that the junction should be considered as rigid objects that can be boosted along the giant directions. Their mass does not depend on the angular momentum that the
 junction carries. We also discuss new features about conditions of marginal stability that arise in these configurations due to the more general geometric structure of LLM geometries. We also point out that one can construct more general BPS networks in these setups. Finally we conclude.

\section{BPS $(p,q)$ Open Strings (Segments)}\label{sec:seg}

As is well known, fundamental strings can end on D-branes. D3-branes are special in that they are invariant under S-duality of the type IIB string. Using this information one knows that D-strings can also end on D3-branes, as well as any $(p,q)$ string bound state. These are obtained from the orbit of the fundamental string
under the $SL(2,{\mathbb Z})$ S-duality transformations.

First, we consider open magnon solutions for open $(p,q)$ strings ending on a pair of giant gravitons wrapping an $S^3$ inside either $S^5$ or $AdS_5$. These are described by the same Nambu-Goto action as the fundamental string, up to a prefactor corresponding to the string tension
\begin{equation}
    S_{(p,q)}= \frac{\sqrt{\lambda}}{2\pi}|p+\tau q| \int d^2 \sigma \sqrt{-\det\gamma_{ab}}.
\end{equation}
Solutions of these equations that are known for the fundamental string are also solutions for the $(p,q)$ string, just with a different tension. The quantity $\sqrt{\lambda}$ is the fundamental string tension and $\tau$ is the complex axion-dilaton coupling constant of type $IIB$ string theory.
These  are BPS with respect to the central charge extension of the symmetry algebra on the worldsheet of the string \cite{Beisert:2005tm}.

A natural choice of coordinates for this problem are the half-BPS coordinates of \cite{Lin:2004nb}. In these coordinates, the geometry is split naturally into two regions in a plane (the LLM plane): the inside and outside of a droplet of radius $r_{AdS}$. This is actually the eigenvalue structure of matrix quantum  mechanics of the half BPS states themselves \cite{Berenstein:2004kk}.

We will work in units where $r_{AdS}=1$ for simplicity. The energy of such sigma model configurations is given by \cite{Berenstein:2020grg, Berenstein:2020jen}:
\begin{equation}\label{dispersion}
    \mathcal{E}= \Delta-J = \sqrt{Q^2+ \frac{\lambda}{4\pi^2}|p+q\tau|^2 |\mathcal{Z}|^2},
\end{equation}
no matter if the strings end inside the  LLM droplet (sphere giant gravitons) or outside (AdS giants). The essence of the paper \cite{Berenstein:2020jen} is that the sigma model solutions can be built in any LLM geometry, not just global $AdS_5\times S^5$ and the formulae look the same. By extension of these ideas, what we're presenting here can be performed in any LLM geometry with extra giant gravitons as well, without any change in the basic formulation of the problem.

The sigma model solutions for both types of endings on sphere giant gravitons or AdS  giants are very similar.
Here $J$ is the angular momentum associated to the LLM geometry (the half-BPS angular momentum of the giant gravitons plus strings), $Q$ is the angular momentum along the directions of the worldvolume of the giant graviton, which is only carried by the strings and $\mathcal{Z}= \xi-\Tilde{\xi}$ is the length of the string in these coordinates. For giant gravitons $Q$ is along the $SO(4)$ unbroken R-charge directions. For AdS giants $Q$ is a momentum along the $S^3$ of $AdS_5$.
The difference between configurations ending on $AdS$ or $S^5$ giants rather than simple giant magnons comes from the condition that $J$ remains finite for the piece of the string itself. For $AdS$ giants  the central charges $\xi, \Tilde{\xi}$ lie outside the unit disk. Geometrically, these correspond to the radial position of the AdS giant gravitons inside of $AdS_5$.
\begin{equation}
    \xi_{\text{out}} = e^{i \theta}\cosh \rho,
\end{equation}
with $\rho$ being the usual radial coordinate  of global $AdS_5$, and $\theta$ is the coordinate along the equator of $S^5$. When the magnitude of both $\xi$ and $\Tilde{\xi}$ are taken to be large, the size of the LLM  droplet becomes irrelevant, so that $\xi, \Tilde{\xi}$ become identified with the Coulomb branch central charge of $\mathcal{N}=4$ super Yang-Mills in flat space \cite{Berenstein:2014zxa}. The central charge for $S^5$ giant gravitons serves as analytic continuation of this Coulomb branch central charge for values inside the droplet in the LLM plane,
\begin{equation}
    \xi_{\text{in}}= e^{i \theta} \cos{\chi},
\end{equation}
where $\chi$ is now the zenith angle in the $S^5$ factor. The important point is that the supersymmetry algebra fixes the kinematics for both kinds of open boundary conditions. All of the relevant quantum numbers are controlled by the central charge extension. These are the conditions for having short multiplets in the ground state of the string. These ensure that the number of polarizations matches the correct number of polarizations for $W$ bosons between two very nearby parallel D3-branes.
The central extension for D-strings (monopoles) is a property of both the field theory and supergravity on flat space, so it is a robust feature when we consider AdS giants.  The central charge is also a property of the spin chain at any value of the coupling constant \cite{Beisert:2005tm} and therefore is also a property of the sigma model of the string (the supersymmetrization of the Nambu-Goto action). In that sense, we have complete control of these states, even when we analytically continue these notions of AdS giant gravitons to sphere giant gravitons. We are leveraging the perturbative field theory knowledge of W-bosons, including the spectrum of monopoles as solutions in ${\cal N}=4 $ SYM.  Their masses are controlled by central charges \cite{Witten:1978mh}. We are also using
the existence of the central charge extension in both the spin chain and by extension the  supergravity sigma model to infer properties of these states. Moreover, we are utilizing S-duality. Basically, with these chains of reasoning we just need to do minimal replacements on what we already know for the fundamental string to obtain the information for all $(p,q)$ strings.

\section{String Junctions}\label{sec:junctions}
We can now consider a string junction configuration consisting of three string segments, each ending on a different giant graviton, such that they are consistent in flat space: they are a member of the  orbit of the $(1,0),(0,1), (1,1)$ string junctions under $SL(2,{\mathbb Z})$ S-duality symmetry.
As before, we assume that the energy contribution of each segment has the form \eqref{dispersion},
\begin{equation}
    \Delta-J= \mathcal{E} =\sum_{i=1}^3 \sqrt{Q_i^2+ T_i^2 |\mathcal{Z}_i|^2},
\end{equation}
Where $T_i$ is the tension of each of the segments, and $|\mathcal{Z}_i|$ their length. There will be three values of $z_{1,2,3}$ corresponding to the giant graviton endpoints, and a further common value $z$ where the strings meet. In this construction $|\mathcal{Z}_i|=|z_i-z|$ is well defined geometrically.  Since string junctions are BPS in flat space,
we expect that there is no extra energy from the junction itself. Just the energy associated to the  geometry of the different strings joining.

A priori, such a configuration with fixed $Q_i$ has a different angular velocity for each string piece along the giant graviton directions. This would cause it to split into pieces that don't  end
consistently.
To prevent this from happening, dynamical forces at the junction would act so as to share the angular momentum and force the system to move together.
We should look for configurations where the total angular momentum of the system is fixed, $Q=\sum_{i=1}^3 Q_i $ and is allowed to be shared. What that means is that the $Q_i$ are not fixed initially, but should be fixed dynamically.

Now that we have a proposed string junction configuration, if it is going to be BPS it will need to be a minimum of the energy $\Delta-J$ at fixed $Q$. We need to vary over $z$ and the $Q_i$, but we need to keep $Q$ fixed since it is a constant of motion (the angular momentum of the junction in the directions of the worldvolume of the giant gravitons).

To find the energy of such a bound state, we can minimize the following  Hamiltonian, with the total angular momentum being fixed by a Lagrange multiplier
\begin{equation}
    \mathcal{H}= \mathcal{E}+ \beta\left(Q- Q_1-Q_2-Q_3\right)
\end{equation}
Clearly, varying the energy with respect to $\beta$ fixes the total angular momentum:
\begin{equation}
    Q=Q_1+Q_2+Q_3.
\end{equation}
Actually, the Lagrange multiplier $\beta$ has a nice physical interpretation as the angular velocity of the string segments.  Minimizing with respect to $Q_i$ gives
\begin{equation}
\beta = \frac{\partial E_i}{\partial Q_i}\label{eq:angvel}
\end{equation}
which shows that $\beta$ is an angular velocity of the string segment conjugate to $Q$. Since $\beta$ is common to all,  the whole string network co-rotates at the same angular velocity in order to keep the junction from splitting.
We proceed now by keeping $\beta$ as a variable and eliminating the $Q_i$.

 This means that the whole system behaves as a single rigid object, rather than as three different string segments. We can then eliminate each of the $Q_i$ in terms of the corresponding central charge and angular velocity :
\begin{equation}
    \frac{\partial \mathcal{H}}{\partial Q_i}=0 \Rightarrow  Q_i= \frac{ T_i |\mathcal{Z}_i| \beta}{\sqrt{1-\beta^2}}
\end{equation}
where we now see a familiar relativistic energy. Here $\beta$ is like a regular velocity relative to the speed of light and $ T_i |\mathcal{Z}_i|$ is the mass of the particle.
The velocity $\beta=1$ is the speed of light in the giant graviton sphere direction (either in AdS or in the sphere).

After imposing the constraint on the Hamiltonian, the energy can be expressed as a function of the angular velocity $\beta$ and the masses $T_i |\mathcal{Z}_i|$:
\begin{equation}
    \mathcal{E}= \frac{1}{\sqrt{1-\beta^2}}\times \sum_{i=1}^3 T_i |\mathcal{Z}_i|.
\end{equation}
Finally, the angular velocity $\beta$ can be eliminated in terms of the total angular momentum $Q$ by solving the constraint equation,
\begin{equation}
    Q= \frac{\beta}{\sqrt{1-\beta^2}}\sum_{i=1}^3 T_i |\mathcal{Z}_i|.
\end{equation}
As expected, all of the quantum numbers of the system are related to the net effective mass at zero velocity by simple relativistic kinematic factors.
Substituting these conditions into the Hamiltonian tells us that energy of the system follows the dispersion relation of a centrally extended BPS state:
\begin{equation}
\begin{aligned}
     \mathcal{E}&= \sqrt{Q^2 + M_{\text{total}}^2}  \\
     M_{\text{total}}&= \sum_{i=1}^3 T_i| \mathcal{Z}_i|
\end{aligned}
\end{equation}
This is similar to what occurs when one considers bound states of giant magnons as in \cite{Dorey:2006dq}, where the total ``mass" is controlled by the net central charge of the magnon partons and the global momentum of the bound state.

The problem of minimizing an expression of the form
\begin{equation}
M_{\text{total}}= \sum_{i=1}^3 T_i| \mathcal{Z}_i| \label{eq:sameasflat}
\end{equation}
is the same problem that appears in flat space D-brane setups if the $z_i$ are positions of parallel D3-branes in flat space. That is exactly the same problem that is solved in field theory and $(p,q)$ strings suspended between D-branes in \cite{Bergman:1997yw}. We assume that this problem has a well known solution. This also means that the states we find carry over to the correct $(p,q)$ junction  states in the Coulomb branch limit described in \cite{Berenstein:2014zxa}.

 The problem we find after this setup becomes independent of the angular momentum carried by the configuration, namely  $Q$. This means that the junction is not deformed in shape (or mass) by the motion in $AdS$. It should be thought of as a rigid object.

Once the energy is minimized with respect to the position of the junction, the corresponding Hamilton equation says that the net force on the junction vanishes:
\begin{equation}
    \delta \mathcal{E}= \frac{1}{\sqrt{1-\beta^2}}\times \sum_{i=1}^3 T_i \frac{\delta \Bar{\mathcal{Z}_i}\mathcal{Z}_i}{|\mathcal{Z}_i|} + c.c.=0.
\end{equation}
When the position of the junction is chosen to satisfy this equation of motion, the sum of the tension unit vectors $T_i\frac{\mathcal{Z}_i}{|\mathcal{Z}_i|}$ must vanish. This condition is identical to the condition on flat-space BPS string junctions up to the kinematic Lorentz factor $\gamma= \frac{1}{\sqrt{1-\beta^2}}$, which just corresponds to a boost of the original  configuration.

This gives a geometric construction of the solution, as the angles of the segments at the junction are determined from these equations. The loci of points for a third vertex, where the junction lies, to a fixed segment (the segment between two giants) is a circle. The correct position of the junction, if it exists,  is the common intersection point of three circles determined this way.

We should point out that these do not exist on their own. The strings are ending on compact D-branes and carry charge with respect to the worldvolume of the D-brane. This charge must be cancelled by other strings ending on the same D-brane with the opposite orientation on the endpoint.
For fundamental strings, these features can be seen in the combinatorial construction of the fundamental string states states explicitly (see \cite{Balasubramanian:2004nb, deCarvalho:2020pdp}). It would be nice if there were a similar construction for monopole charged strings directly from field theory.
In the Coulomb branch limit, they do exist: they are regular dyons of the Yang-Mills Higgs field. What we mean here is to have more details of the monopole charged strings ending on sphere giant gravitons that we are discussing here.

\subsection{Lines of Marginal Stability}

It is well known that the existence or absence of certain string junctions is controlled by lines of marginal stability that depend on the position of the branes
 and the coupling constant (see \cite{Bergman:1997yw} for some examples). The basic problem of marginal stability will to a first approximation look the same as in flat space, because equation \eqref{eq:sameasflat} is the same as that of a flat space problem. The existence or not of BPS solutions will then be determined by the lines of marginal stability that are known.

There is an  important difference however. The states corresponding to these string junctions are normalizable (have finite energy) only whenever each string segment stays entirely inside or outside a given droplet in the LLM plane \cite{Berenstein:2020grg}, which is different than the more naive expectation for existence  of states \cite{deMelloKoch:2015uwu} based on segments between branes or as just magnons in the infinite spin chain for a droplet configuration.
These constraints lead to additional configurations (lines) of marginal stability where the BPS objects can decay or disappear.
The naive decay which also occurs in flat space is into two strings, let us say the $(1,0)$ and  $(0,1)$ string, which can then fly apart and separate. With the extra conditions, the BPS states go to ``infinity'' when any segment of the string junction is becoming tangent to a droplet interface. This can occur at fixed position of the giant gravitons when we change the coupling constant. That change will deform the shape of the junction and can lead to a segment touching the edge of a droplet.  The quantum number that goes to infinity is the amount of momentum along the BPS charge  that is carried by the background D-branes (the quantum number $J$ in $\Delta-J$) that is carried by the junction configuration itself. The quantity
$J$ is the total $J=J_{brane}+J_{junction}$. This does not show up in $\Delta-J$. However, it can be seen explicitly in spin chain computations \cite{Berenstein:2005fa,Berenstein:2006qk} or sigma model computations where the droplet configuration preserves an additional $SO(2)$ symmetry \cite{Berenstein:2020grg}. These are implicit also in the giant magnon geometry itself
\cite{Hofman:2006xt}.

\subsection{BPS Networks}
This class of string junctions are invariant under one-eight of the $32$ supersymmetries of $AdS_5\times S^5$. This is because the junction behaves as a $\frac{1}{4}$-BPS monopole-dyon bound state, while the boundary conditions break this further to one-eight if the giant gravitons are taken to be parallel to each other and transverse to the string network.  On the gauge theory side, this corresponds to a generic BPS configurations on $\mathbf{R}\times S^3$ where all three complex scalar fields have a non-zero vev. Turning on the vevs for all three fields while keeping the D3-giants parallel is only possible for D3 branes that wrap the same $S^3$ inside of the $AdS_5$ factor, while they rotate along different axes in the $S^5$. This is manifest in the classical string picture, where the geometry can be written as a cone over a 6d K\"{a}hler base for a generic eight-BPS configuration \cite{Chen:2007du}.  This base is the higher dimensional analog of the LLM plane, and contains a five-sphere locus of degeneration where the $S^3$ factor shrinks to zero size. Whenever the giant gravitons preserve at least a quarter of the supersymmetries, the geometry can be written in such a way that the radial direction can be continued beyond the coordinate singularity at $r=1$, which restricts the allowed configuration of end points to those that are $\frac{1}{4}$-BPS.  In that case, the giant gravitons can be taken to wrap more complicated holomorphic cycles in a complex unit ball.

In particular, as long as the string boundaries are taken as such, many co-planar junctions can be combined into a single BPS network. As in flat space, these must preserve the same set of supersymmetries as a single junction which means that string networks are also stable configurations as long as they remain completely inside or outside a single  droplet region.

\section{Conclusion}

We considered classical BPS string junctions solutions in $AdS_5\times S^5$, and showed that they satisfy the dispersion relation expected for a BPS state of a centrally extended supersymmetry algebra. This superymmetry algebra is roughly speaking the residual light-cone symmetry of the rotating giant graviton configuration, with the central charges being associated to the positions of the branes in the LLM plane, or its higher dimensional analog. The analysis is qualitatively similar to the situation in flat space, except that the energy of the state receives simple kinematic corrections. These are just from motion along the giant gravitons that they are suspended from.
Apart from that, they behave as rigid objects. Since the unbroken supersymmetries play an important role in controlling the kinematics, one should expect that these BPS junction configuration exists in a generic half BPS geometry, where more than one droplet is used. One way to see this is that there is always a plane that contains the center of the droplet, one of the giant gravitons, and the junction point, and when restricted to this plane the resulting geometry looks $\frac{1}{2}$-BPS, and hence there will be a straight line connecting the junction and the giant graviton. Since this can be done for each section of the junction, this means that such a configuration will satisfy the equations of motion for the Nambu-Goto action. It is possible that these can also be constructed in ${\cal N}=2$ superconformal field theories in the Coulomb branch. They are similar in that only singlets under the $SU(2)$ R-symmetry of the ${\cal N}=2$ would be turned on and there is a central charge on the worldvolume of the strings \cite{Gadde:2010ku,Liendo:2011xb}. The problem of the different strings ending on giants in that type of geometry has not been addressed yet in the literature. Those geometries have conical singularities of type ${\mathbb C}^2/\Gamma$ on the sphere, which would intersect the worldvolume of the sphere giants.
The particular case of  $\Gamma ={\mathbb Z}_n$ has extra symmetry that might make it easier to analyze.
The moduli that control the different coupling constants of the spin chain are modes attached to the singularities \cite{Gukov:1998kk}. This suggests that monodromies of strings around these singularities might have a clue about how to solve the problem (this is similar to  how monodromies of D3-branes can detect the presence of discrete torsion \cite{Berenstein:2000hy}). The AdS fractional brane giants will be stuck at the singularity and their physics is more similar to the Seiberg-Witten theory \cite{Seiberg:1994rs} than to ${\cal N}=4 $ SYM. That is, we expect quantum corrections to the BPS masses on the moduli space. The sphere giants we need are smooth branes with shape $S^3/\Gamma$. it is expected that  ${\cal N}=2$ structure at higher loop orders might help to elucidate how this works in detail. We would like to highlight that there is a program where some of these computation have been done in in the closed spin chain setup \cite{Liendo:2011xb,Pomoni:2013poa,
Pomoni:2019oib,Pomoni:2021pbj}.

From the point of view of the ${\cal N}=4$ spin chain, the fact that the fundamental string can end on a D1-string and preserve supersymmetry suggests that one can describe the supersymmetric boundary conditions that they give rise to directly on the spin chain. Importantly, the ground state of the string will carry a fixed amount of angular
momentum $Q_s$, because its velocity $\beta$ 	will be fixed (this is the content of equation  \eqref{eq:angvel} in this paper). It would be worthwhile to understand the precise decomposition of zero modes, since this can be compared with the low energy effective theory on the worldsheet of the giant gravitons. This information encodes the corresponding boundary conditions that one would expect for the open spin chain arising from $\mathcal{N}=4$ Yang-Mills \cite{Berenstein:2005vf, Berenstein:2006qk,Holguin:2021qes}. In principle, one should be able to reproduce the position of the junctions and the lines of instability from a spin chain computation along the lines of \cite{Vazquez:2006id} (see also \cite{deMelloKoch:2007rqf, deMelloKoch:2007nbd, Bekker:2007ea} and to higher loop orders \cite{Berenstein:2014isa}).

 The same applies to a generic string network; the end points of each segment of $(p,q)$ string will preserve one-half of the spacetime supersymmetries, so there will be a straight string that connects those points as long at this line doesn't cross over into a forbidden region of the LLM  plane. This produces additional situations where BPS disappear not because they cross a line of marginal stability, but because a quantum number in the string junction configuration goes to infinity.

\acknowledgments

The work of D.B. is supported in part by the Department of Energy under grant DE-SC 0011702.

\end{document}